\documentclass[fleqn,twoside,11pt]{article}

\usepackage[numbers]{natbib}
\usepackage{left_eq}   
\usepackage{CSPM}  

\usepackage{pstricks, pst-plot}	
\usepackage{graphicx} 
\usepackage{wrapfig}  
\usepackage[figuresright]{rotating}
\usepackage{subcaption}
\usepackage{float}

\usepackage{amsmath}
\usepackage{amssymb}
\usepackage{bbm}

\def\tr{\mathrm{\,tr}}

\def\i{\textrm{i}}

\def\Z{\mathbb Z}

\def\id{{\mathbbm{1}}}

\usepackage[english]{babel}
\usepackage{blindtext}

\title{Measurments-induced quantum phase transitions}
\author[a,*]{Dragi Karevski}
\author[a]{Michele Coppola}
\author[b]{Emanuele Tirrito}
\author[b]{Mario Collura}
\affil[a]{Universit\'e de Lorraine, CNRS, LPCT, F-54000 Nancy, France}
\affil[b]{SISSA, Via Bonomea 265, 34136 Trieste, Italy}
\email{*dragi.karevski@univ-lorraine.fr}

 \HeaderTitle{Measurments-induced quantum phase transitions}
 \HeaderAuthor{D. Karevski et al.}

\begin{document}
\maketitle             

\thispagestyle{empty}  

\begin{changemargin}{1.5cm}{1.5cm} 
\begin{abstract}
  \small{Dynamical phase transitions induced by local projective measurements
have attracted a lot of attention in the past few years. It has been in
particular argued that measurements may induce an abrupt change in
the scaling law of the bipartite entanglement entropy. In this work we
show that local projective measurements on a one-dimensional quadratic
fermionic system induce a qualitative modification of the time growth of
the entanglement entropy, changing from linear to logarithmic. However,
in the stationary regime, the logarithmic behavior of the entanglement
entropy does not survive in the thermodynamic limit and, for any finite
value of the measurement rate, we numerically show the existence of a
single area-law phase for the entanglement entropy. We give analytical
arguments supporting our conclusions.
  } 
\end{abstract}

\end{changemargin} 


\section{Introduction}

The dynamics of isolated quantum many-body systems undergoing a more or less sudden quench of a global or local parameter 
has been studied extensively this last two decades \cite{Calabrese2005,Amico,Eisler2007,Eisler2008}. 
When the changes of the Hamiltonian parameters are inhomogeneous, varying locally in space and time according to power laws, interesting phenomena may occur close to a quantum critical point where the gap of the Hamiltonian closes and leads to a critical slowing down \cite{Collura2010,Collura2011,Scopa2017,Scopa2018}.  
One important way of characterizing the non-equilibrium properties of the quenched system is through the time evolution of its entanglement which spreads all over the system \cite{Eisert2010,Laflorencie}. 
For short-range interactions, starting from a completely disentangled initial state, typically the spreading of quantum correlations is ballistic, governed by  a Lieb-Robinson bound \cite{LiebRobinson} and leads to  light-cones effects where correlation fronts propagate at a maximum velocity. The easiest way of understanding this is through the so-called quasi-particle picture in integrable one-dimensional systems. Indeed, in those systems just after the quench infinite-life time quasi-particle excitations pairs are generated at each points and start to 
propagate ballistically in opposite directions, building up entanglement between regions that are far appart  \cite{Calabrese2005,Alba2018}. 
At long times, the system relaxes asymptotically to a local Generalized Gibbs state which is specified by an infinite set of conserved quantities which  reflects the extensive nature of the entanglement  \cite{Rigol2008,Ilievski2015,Ilievski2016,Abanin2019}.

However, many factors may affect the nonequilibrium spreading of the entanglement and
its  scaling behavior could vary in out-of-equilibrium monitoring.  
On these lines, dynamical phase transitions induced by local projective measurements have attracted a lot of attention in the last few years. This has started with the exploration of the entanglement properties of quantum circuits made of random unitaries alternating with local projective measurements, see \cite{Nahum2017,Nahum2019},  \cite{Li2018} for seminal works which soon have been followed by many others, see \cite{Chan2019,Vasseur2019,Bao2020,Choi2020,Gullans2020,Jian2020,Zabalo2020}.
It has been shown that by increasing the rate of the local measurements the bipartite entanglement entropy, which is a measure of entanglement, switches from the volume law $S\sim \ell^d$, where $\ell$ is a typical scale of the reduced system and $d$ the space dimension of the system, to the area law $S\sim \ell^{d-1}$ which is in general associated to many-body ground states properties or to many-body localized states. 
A similar transition was reported on the quantum stochastic trajectories of many body systems subjected to local continuous monitoring \cite{Cao2019,Alberton2021,Fuji2020,Coppola2022} through numerical studies and analytical means for one dimensional free-fermion systems.
The transition is driven by the competition between the growth of quantum entanglement under unitary time evolution and the disentanglement induced by the projection of the system state on eigenstates of local operators under projective measurements. 

In this proceedings paper we report on the dynamical behavior of the entanglement entropy of a simple free-fermionic one-dimensional system under local projective measurements \cite{Coppola2022}. 
In this extended simple model we study in particular how the bipartite entanglement
entropy is affected by the monitoring of the local particle density. As a main result, we find that the volume-law phase is
absent for any measurement rate. During the initial  transient regim,
we observe that any finite measurement rate induces an abrupt change in the entanglement dynamics, from linear to logarithmic growth. Moreover,  numerical evidences show 
that the average of the asymptotic time entanglement entropy undergoes
a  transition from the volume-law to the area-law phase for
any measurement rate in the thermodynamic limit. 
Logarithmic scaling in the steady state is thus a finite size effect. 
The next section specifies the model we consider and section 3 is devoted to our results. Finally we give some conclusions and discussion. 

\section{Hopping fermions under measurements of the local density }
\subsection{Hopping Hamiltonian}
One of the most simple situation that realizes the measurement-induced  transition discussed so far is the case of free fermions on a one-dimensional lattice with unitary evolution generated by the quadratic hamiltonian (with periodic boundary conditions)
\begin{equation}
H= -\frac{1}{2}\sum_{i=0}^{L-1}\left(  c_i^\dagger c_{i+1} + c_{i+1}^\dagger c_i\right)  \; , 
\label{H1}
\end{equation}
where $L$, which is supposed to be even, is the size of the ring and where the $c_i$ and $c_i^\dagger$ are the fermi annihilation and creation operators satisfying the canonical anticommutation relations 
$\{ c^\dagger_i, c_j\}= \delta_{ij}$, $\{c_i,c_j\} =  \{c^\dagger_i,c^\dagger_j\} = 0$.
 This model is readily diagonalized by the introduction of the fermionic Fourier modes 
$\eta_k = \frac{1}{\sqrt{L}} \sum_{j=0}^{L-1} e^{-\textrm(i) 2\pi kj/L} \, c_j$
and the hermitic conjugates $\eta^\dagger_k$. The diagonal Hamiltonian is thus
$H= \sum_{k=-L/2}^{L/2} \epsilon_k \eta^\dagger_k \eta_k$ 
with the single particle excitation energy $\epsilon_k = -\cos (2\pi k/L)$. 
The Hamiltonian (\ref{H1}) commutes with the total
number of particles which is given by $\hat{N} = \sum_j c^\dagger_j c_j$ or in Fourier space by  $ \sum_k\eta_k^\dagger\eta_k$.  
Thanks to the quadratic structure of the Hamiltonian, the unitary dynamics generated by $e^{-\i tH } $ preserves the Gaussianity of an initial Gaussian state $\rho(0) \propto e^{-\sum_{ij} c^\dagger_iT_{ij}  c_j}$. For such Gaussian state, Wick theorem applies and one may specify completely the state of the system by its two-point correlation matrix $C$ with entries
\begin{equation}
C_{ij} = \tr\{c^\dagger_i c_j \rho\} \; . 
\end{equation}
At a later time, under unitary dynamics the two-point correlation matrix $C$ evolves according to 
$C(t+s ) = R^\dagger(s) C(t) R(s)$
where $R(s)$ is a unitary matrix with elements
$R_{ij}(s) = \frac{1}{L} \sum_{k=-L/2}^{L/2} e^{-\i 2\pi (i-j)k/L} e^{- \i \epsilon_k s} $.
In the thermodynamical limit $L\rightarrow \infty$ one can show that $R_{ij}(s) \sim \i^{i-j} J_{i-j}(s)$ where $J_n(s)$ is a Bessel function of the first kind. This is all we need to characterize completely the unitary dynamics of such a system.

\subsection{Measurement protocol}
The system is initially in a Gaussian state, which is not a stationary (eigenstate)  state,     with typically very short-ranged correlations, that is close to a product state $|\Psi(0)\rangle \simeq \otimes_j |\phi_j\rangle$. 
In the subsequent dynamics the unitary evolution generated by $H$ is perturbed by random interactions with local measuring apparatus that project locally the state to an eigenstate of the measured observables, according to Born rule. 
To be more precise, consider a local observable $Q_\Omega$, defined on a compact support $\Omega \subset \Z$, 
\begin{equation}
Q_{\Omega} = \sum_p q_p P^{(p)}_\Omega \; , \qquad \sum_p P^{(p)}_\Omega = \id_\Omega \; ,
\end{equation}
where the $P^{(p)}_j$ are  orthogonal projectors  associated to the eigenvalue $q_p$ of  $Q_\Omega$ on the corresponding subspace.
Immediately  after the measurement of the local observable $Q_\Omega$, with  outcome $q_k$, the state is projected according to 
\begin{equation}
|\Psi \rangle \longrightarrow \frac{ P^{k}_\Omega |\Psi\rangle }{\langle \Psi | P^{k}_\Omega |\Psi\rangle} \; . 
\label{proj}
\end{equation}
In general, projective measurements are not preserving the Gaussianity of the state. However, 
if we consider only the measurements of  local particle densities, that is  $\hat{n}_j = c^\dagger_j c_j$,  the non-unitary dynamics remains Gaussian preserving. 
To see this, notice that the local density operator can be represented as $\hat{n}_j =  1. P_j^{(1)} + 0. P_j^{(0)}$ where $P_j^{(1)} +  P_j^{(0)} = \id_j$ which implies  
$\hat{n}_j =  P_j^{(1)}$ , $ \id_j-\hat{n}_j =   P_j^{(0)} $. 
With the operator identity
$e^{\nu \hat{n}_j } = \id_j + (e^\nu -1) \hat{n}_j $
the projectors $P_j^{(0)}$ and  $P_j^{(1)}$ can be expressed as a limit of Gaussian operators:
$$
\id_j - \hat{n}_j = \lim_{\nu \rightarrow \infty} e^{- \nu \hat{n}_j } \; , \quad 
\hat{n}_j =  \lim_{\nu \rightarrow \infty} \frac{e^{\nu \hat{n}_j }}{e^\nu -1} \; . 
$$
The initial state being  Gaussian, $\rho \propto e^{\sum_{ij}c^\dagger_i M_{ij} c_j} $,  the projection rules associated to either $\id_j - \hat{n}_j $ or $\hat{n}_j $  lead to 
$
e^{\pm \nu \hat{n}_j } e^{\sum_{kl}c^\dagger_k M_{kl} c_l} e^{\pm \nu \hat{n}_j }
$
which, thanks to Baker-Campbell-Hausdorff formula, is still  a Gaussian state
$e^{\sum_{kl}c^\dagger_k K_{kl} c_l}$  with a new coupling matrix $K$. 

The projection rule (\ref{proj}) translates for the two-point correlation matrix, thanks to Wick theorem, into
\begin{equation}
C_{ij}(x,t) \longrightarrow \delta_{ik}\delta_{jk} + C_{ij}(x,t) - \frac{C_{ik}(x,t) C_{kj}(x,t)}{C_{kk}(x,t)} 
\label{Cij}
\end{equation}
if the outcome is $1$ and otherwise
\begin{equation}
C_{ij}(x,t) \longrightarrow -\delta_{ik}\delta_{jk} + C_{ij}(x,t) + \frac{(\delta_{ik} - C_{ik}(x,t)) (\delta_{jk}- C_{kj}(x,t))}{1-C_{kk}(x,t)} \; . 
\label{1-Cij}
\end{equation}
Indeed, just after an outcome $1$ the projection rule transform $C_{ij}= \tr\{c^\dagger_i c_j \rho\}$ into 
$
\tr\{c^\dagger_i c_j \hat{n}_k \rho \hat{n}_k \} = \tr\{\hat{n}_k c^\dagger_i c_j \hat{n}_k \rho  \} = \langle c^\dagger_k c_k c^\dagger_i c_j c^\dagger_k c_k  \rangle
$
and  after normal ordering this six-point correlation function, one is left with
$
\delta_{ik}\delta_{jk} \langle c^\dagger_k c_k\rangle + \langle c^\dagger_k c^\dagger_i c_j c_k\rangle
$
which, with $\langle c^\dagger_k c^\dagger_i c_j c_k\rangle = \langle c^\dagger_k c_k \rangle \langle c^\dagger_i c_j\rangle
-  \langle c^\dagger_k c_j \rangle \langle c^\dagger_i c_k\rangle $ and after dividing by the proper normalization $\tr\{\hat{n}_k \rho \hat{n}_k\} = \tr\{\hat{n}_k \rho \}= \langle \hat{n}_k \rangle = C_{kk}$, leads to the expression (\ref{Cij}). 

Let us recapitulate the dynamics on which the monitored hopping fermionic system is submitted: 
for each infinitesimal time step $dt$ and each site $k$ of the chain, we draw a random number $q_k\in [0,1]$ and if $q_k \le dt/\tau$ the measurement of the local occupancy $\hat{n}_k$ is performed, otherwise the evolution is unitary. Consequently, $1/\tau$ defines the rate at which the measurements are performed. When such a measurement occurs, we draw an other random number $p_k$ from the uniform distribution $[0,1]$ and if $p_k \le C_{kk}(t) =  \langle \hat{n}_k(t) \rangle$ then the correlation matrix is transformed according to (\ref{Cij}) and according to (\ref{1-Cij}) otherwise. 

\begin{figure}[H]
\centering
 \includegraphics[width=0.6\textwidth]{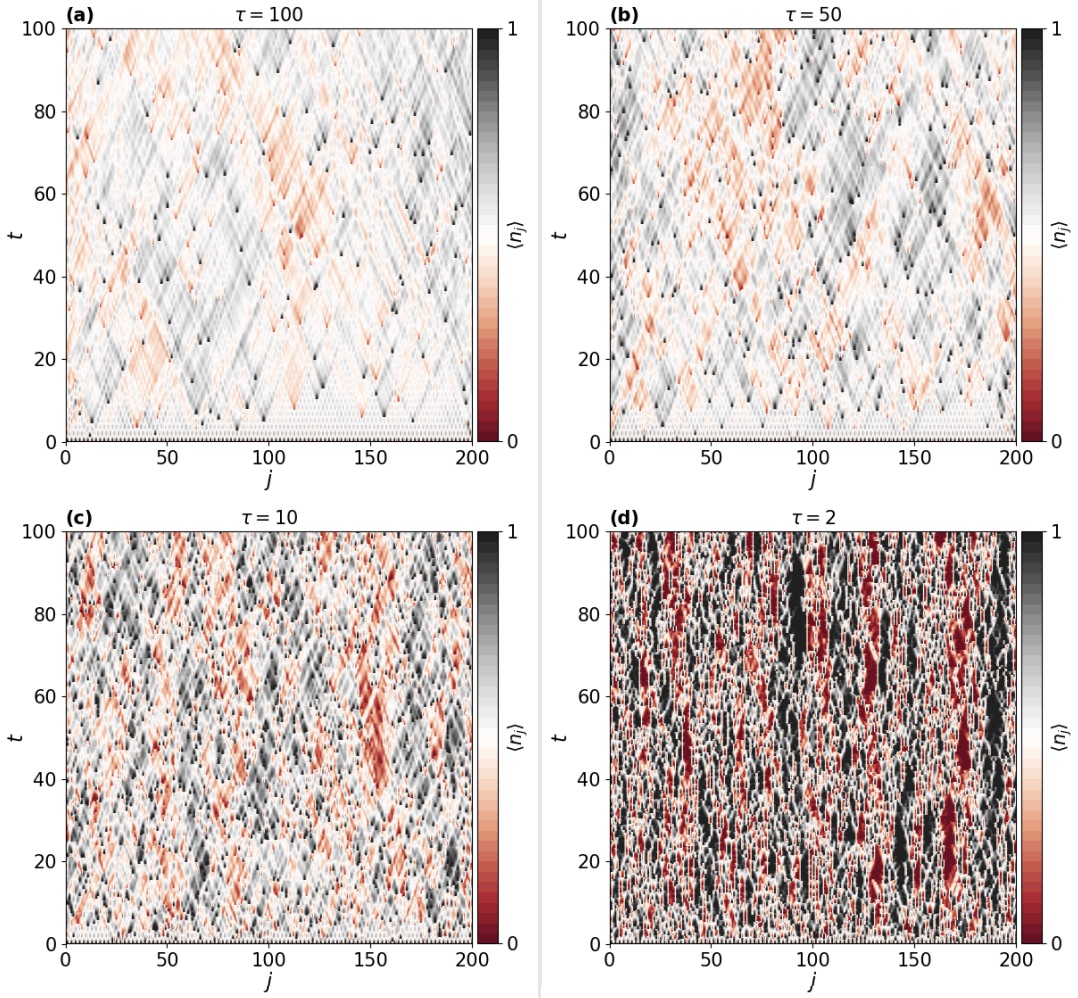}
 \caption{Evolution of the particle density after a sudden quench from a N\'eel state. The different panels represent a typical trajectory where random projective measurements of the local occupation number occur with different rates $1/\tau$. }
 \label{fig-PD1}
\end{figure}

Following the state of the system all along the measurements outcomes, one gets a (random) quantum trajectory of the system. 
In figure~\ref{fig-PD1} we show snapshots of the typical evolution starting from a N\'eel product state $|\Psi(0)\rangle = |1,0,1,\dots, 0\rangle$  (which is a Gaussian state) for several measurements rates. We observe that without any measurement the time evolution follows the usual unitary dynamics and the system melts, relaxing toward the infinite temperature state.   

When the local measurements are very dilute, for $1/\tau \ll 1$, we observe the appearance just after the measurement of  localized  spikes on top of the infinite temperature landscape. These localized excitations spread then ballistically and relaxe algebraically in time, which affects the whole unitary dynamics even for infinitesimal amount of measurements. 
At very large measurement rates, for $\tau \sim O( 1)$, we observe long vertical stripes reflecting the fact that we are close to the Zeno regime  where the local state is measured again and again, locking its unitary spreading. In this case one expects very short entangled domains. 

\section{Entanglement entropy dynamics}
\subsection{Early dynamics of the entanglement entropy}
For a pure state $|\Psi\rangle $  the  bipartite entanglement entropy between a subsystem $s$ and the rest of the system, the so called environment, is defined as the von Neumann entropy 
$
S = -\tr_S \{\rho_s \ln \rho_s\}
$
of the reduced density matrix 
$
\rho_s=\tr_{Env.}\{ |\Psi\rangle \langle \Psi |\} $.
Notice that even if the total system is a pure state, in general the state of the reduced system is not pure anymore. 
In the hopping fermions system that we consider here, with a dynamics that preserves the Gaussianity of the state, the time-dependent bipartite entanglement entropy, for a subsystem consisting of $\ell$ contiguous lattice sites is given by \cite{Calabrese2005}
\begin{equation}
S_\ell(t) = - \sum_{k=1}^\ell \left[ \lambda_k (t) \ln \lambda_k(t) + (1-\lambda_k) \ln (1-\lambda_k) \right]
\end{equation}
where the $\{\lambda_k\}$s are the eigenvalues of the $\ell\times \ell$ two-point correlation matrix $C|_\ell$ restricted to the subsystem $\ell $, that is with entries $C_{ij}$ for $i,j\in \ell$. 

Without any measurements, the dynamics is unitary and from the initial zero entangled N\'eel state, the entanglement entropy starts to grow linearly in time up to a typical time $t =\ell/2$ since the maximum of the quasi-particule velocities $v_k = \partial_k \epsilon_k$ is one. After that time, the system enter  a regime where the entanglement entropy is saturating toward the extensive value $\ell \ln 2$, leading the so called  volume law (remember here that space dimension  is one). 

In the opposite limite, at infinite rate of measurements, $1/\tau \rightarrow \infty$, the system is permanently measured and the unitary dynamics has no time to spread entanglement: the state remains completely disentangled and the entanglement entropy $S_\ell(t)$ stays very small and this corresponds trivially to an area-law (entanglement entropy of order one independent on $\ell$). 

The fate of the intermediate regime, when the local densities are measured at finite rate, is less clear, even if one expects a lower entanglement than in the unitary situation. The main question that arises is to know if  there is a threshold rate above which the volume-law entanglement switches toward the area-law.  

\begin{figure}[H]
\centering
 \includegraphics[width=0.6\textwidth]{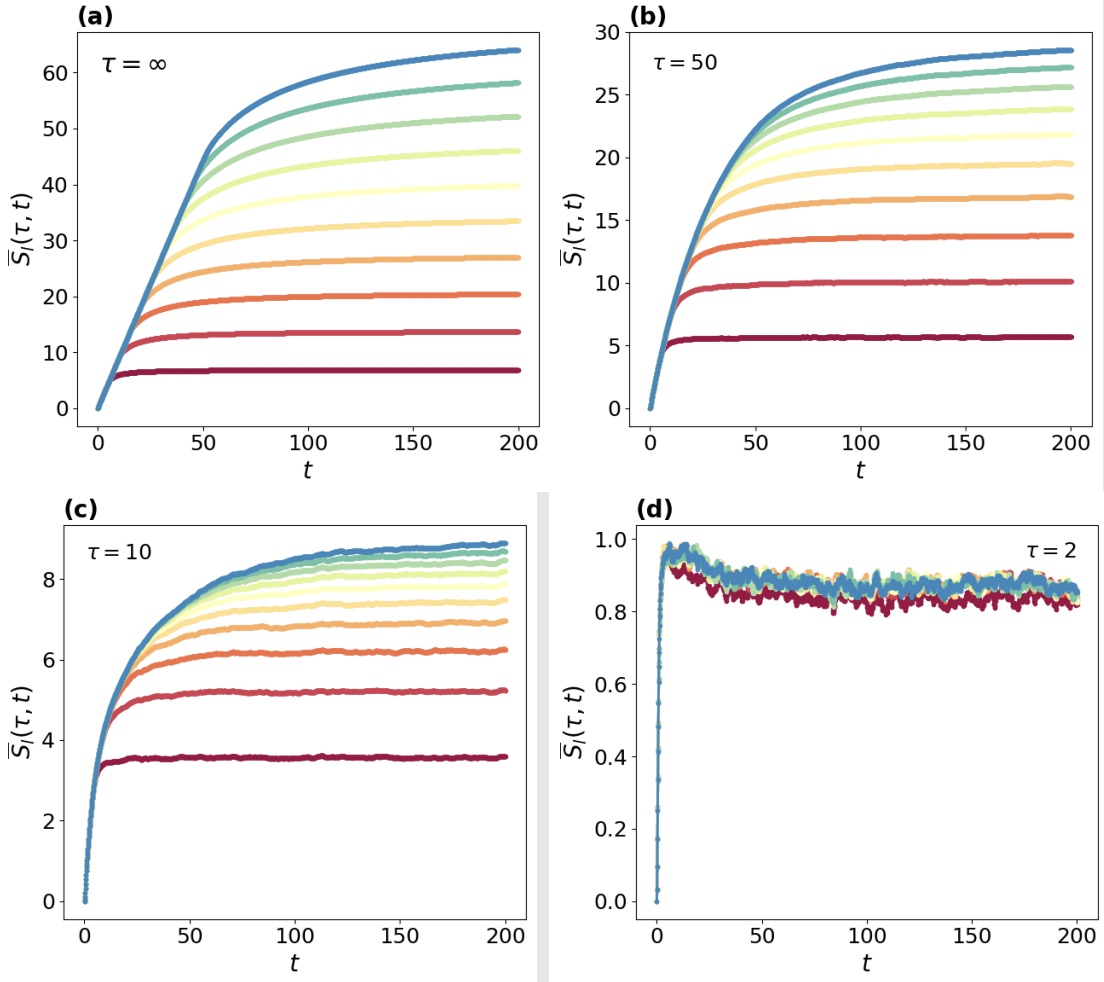}
 \caption{Evolution of the averaged entanglement entropy for  different rates $1/\tau$ of measurements and subsystem sizes $\ell=10, 20, 30, \dots , 100$ from bottom to top. }
 \label{fig-EE1}
\end{figure}

We have analyzed that question by a careful numerical study of the entanglement entropy for different sub-system sizes $\ell$ and averaged over up to a thousand of different quantum trajectories for total system sizes up to $L=400$. We show the results for the time evolution of the averaged entanglement entropy on figure~\ref{fig-EE1}.

The unitary case, $\tau =\infty$, shows clearly the expected initial linear increase followed by the saturation toward the volume law $\ell \ln 2$.  
Increasing the rate of measurements $1/\tau$,  we observe from the numerics that the linear growth
of the entanglement entropy suddenly changes to a logarithmic growth $S_\ell(t) \simeq a(\tau) \ln t +b(\tau)$ (see
Figs. 2(b) and 2(c)), which eventually saturates at large times. 
Finally, at very high measurements rate, in the Zeno limite, see Fig. 2(d), we observe a rapid saturation of the entanglement entropy toward a small size-independent value, the so-called area law.

\subsection{Stationary entanglement}
In the asymptotic time limite (in order to avoid finite size effects we need to stop the simulations at $t_{max}< (L-\ell)/2$) the 
entanglement entropy reaches a constant value which has a non trivial dependence on both the subsystem size $\ell$ and the measurement rate $1/\tau$. In particular, we want to determine if there is a threshold value $\tau^*$ separating an intermediate logarithmic regime, with entropy scaling as $\ln \ell$, from the area-law regime. 

\begin{figure}[H]
\centering
 \includegraphics[width=0.3\textwidth]{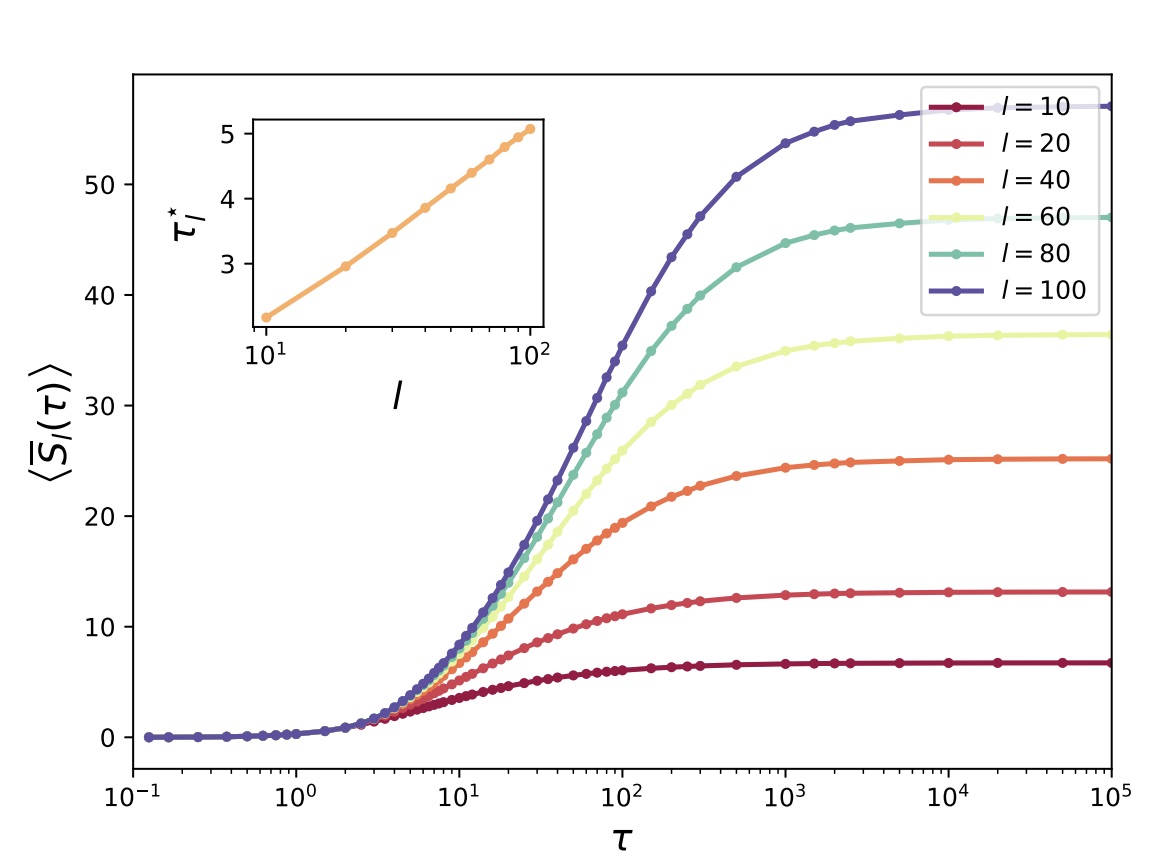}
 \includegraphics[width=0.3\textwidth]{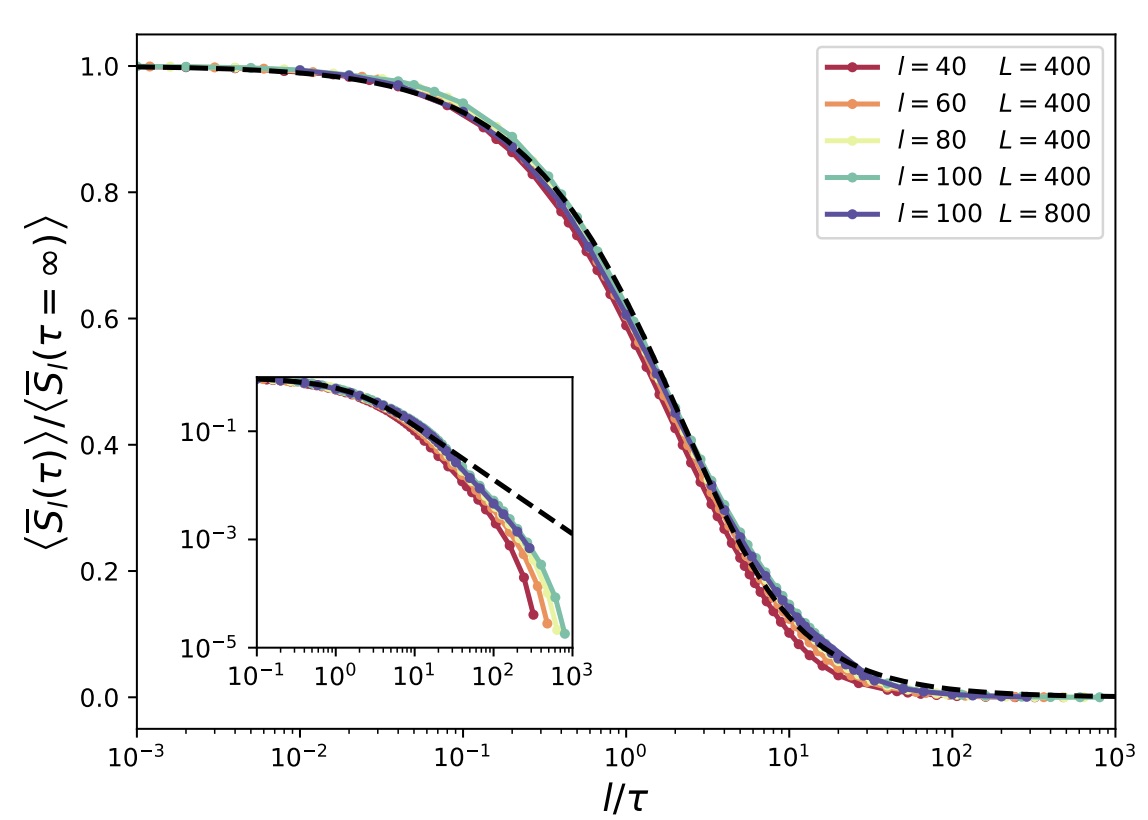} 
 \caption{ a) Stationary entanglement entropy versus $\tau$ for different subsystem sizes $\ell$. The inset shows the divergence of the inflection point $\tau^*$ as a function of $\ell$ (in semi-log scale). b) the same as a) but rescaled by $S_\ell (\tau=\infty)= \ell \ln 2$ and as a function of $\ell/\tau$. 
 }
 \label{fig-EE3}
\end{figure}

We report on figure~\ref{fig-EE3}a our numerical analysis of the steady entanglement entropy. We observe that at  small $\tau$ (very high measurement rate) the steady entanglement entropy is independent on $\ell$. For $\tau > \tau^* \sim  \ln \ell$ on the contrary we observe a clear dependence on the subsystem size $\ell$. 

This behavior is explained by the fact that the measurements introduce a typical correlation length scale which grows exponentially with $\tau$:  $\xi(\tau) \sim e^{\alpha \tau}$. As a consequence, for a subsystem size $\ell \gg \xi(\tau)$ 
only  the sites close to the border of the subsystem are correlated with the
rest of the system which leads the $\ell$-independent area law for the entanglement entropy. On the contrary, as $\xi(\tau)$ gets larger and larger, more and more sites are involved in generating correlations with the
rest of the chain and when $\xi(\tau) > \ell $ the entire subsystem contributes to the entanglement entropy and this essentially results in a volume-law behavior.
As a consequence, in the thermodynamic limit $\ell \gg 1$, at any finite rate of measurements $1/\tau$ the steady entanglement entropy follows an area-law $\propto \ln \xi(\tau)$.   
This is confirmed by analytical arguments based on the quasi-particles picture, see \cite{Coppola2022} for details, which predict that 
the steady entanglement entropy should scale as
$
\frac{S_\ell(\tau, t\rightarrow \infty) }{\ell \ln 2} = f(\frac{\ell}{\tau}) 
$
where the scaling function $f(x)$ has limits  $\lim_{x\ll 1} f(x) = 1$ and  $\lim_{x\gg 1} f(x) = 1/x$. This is shown in figure~\ref{fig-EE3}b. 

\begin{figure}[H]
\centering
 \includegraphics[width=0.3\textwidth]{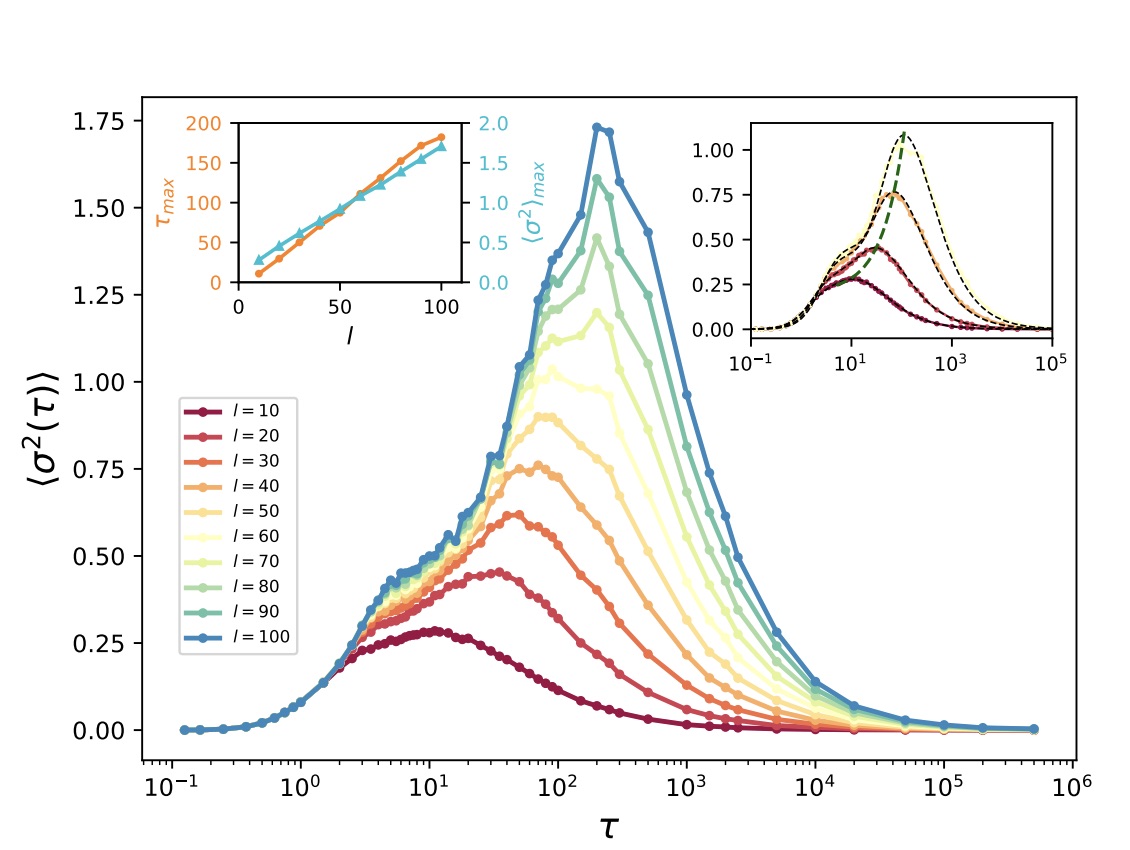}
 \includegraphics[width=0.3\textwidth]{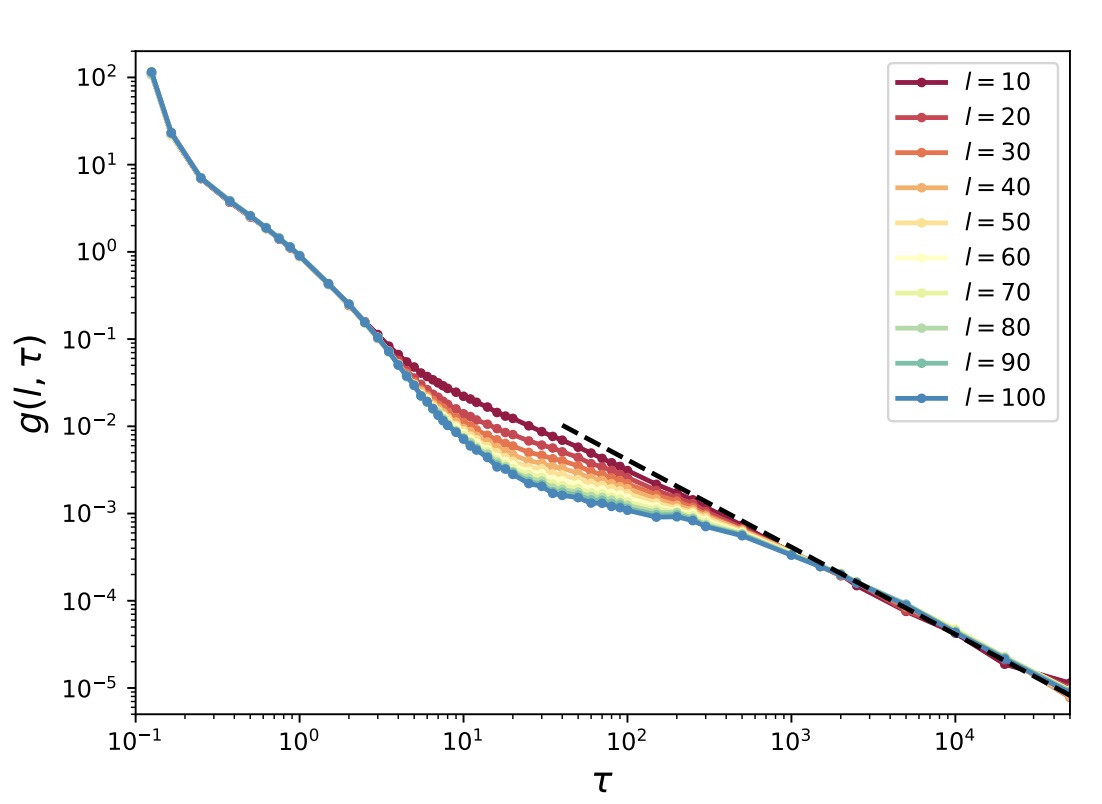} 
 \caption{ a) Fluctuations of the stationary entanglement entropy as a function of $\tau$ for different subsystem sizes $\ell$. 
 The absolute maximum point $\sigma^2_{max}$ and its position $\tau_{max}$  increase linearly with the subsystem size $\ell$ as shown in the inset on the left.
 b)  Scaling of the fluctuations for small measurement rates. The dashed line represents the asymptotic behavior
expected theoretically, see main text for details. }
 \label{fig-FEE1}
\end{figure}

On the same lines we have also considered the fluctuations of  the entanglement entropies over many different quantum trajectories, as measured by the variance of the empirical distribution. Figure~\ref{fig-FEE1} shows our numerical evaluation of the variance of the entanglement entropy as a function of $\tau$ and for different subsystem sizes. 

From the numerical data, we see that  the variance is $\ell$-independent for very high measurement rates and 
we see that it approximately decays as $1/\tau$ for very low measurement rates. This behavior in the extreme cases is not a surprise: at very high rates, we are close to the Zeno regime and then
we do expect that also higher momenta of the entanglement entropy  are size independent. 
At low rates, the behavior may be easily understood noticing that the probability to
have multiple measurement events at each time step $dt$ is approximately zero. Under this assumption, the measurement becomes a Poisson process and from it one can show that the ratio $g(\ell,\tau)= \sigma_\ell^2(\tau) / (S_\ell(\tau))^2$ is essentially $\ell$-independent and proportional to $1/\tau$ at low measurement rates \cite{Coppola2022}. This is shown in figure~\ref{fig-FEE1}b.

\section{Conclusions}
By computing the entanglement entropy of the fermionic system during
its time evolution under projective measurements of its local particle densities, we have found that the entanglement shows
a logarithmic growth in time before reaching a final stationary value. We have shown that the properties of the stationary
entanglement entropy reflects a dynamical transition from a volume-law in the unitary case up to an area-law at any value of the measurement rate, excluding the possibility of an intermediate logarithmic phase at finite measurements rate.

This proceedings paper is based on the work \cite{Coppola2022} and all the figures presented here are extracted from that paper. The Society of Physicists of Macedonia and especially the organizers of the Ohrid conference 2022 are gratefully acknowledged.






\end{document}